\documentclass[preprint]{ptephy_v1}

\usepackage{amsmath,amssymb}
\usepackage{color}
\usepackage{graphicx}

\begin{document}


\title{
Determination of a dineutron correlation in Borromean nuclei via a quasi-free knockout ($p,pn$) reaction
}


\author[1,2]{Yuma Kikuchi}
\affil{Department of Physics, Osaka City University, Osaka 558-8585, Japan \email{yuma@sci.osaka-cu.ac.jp}}
\author[3,1]{Kazuyuki Ogata}
\affil{RIKEN Nishina Center, Wako 351-0198, Japan }
\author[2,4]{Yuki Kubota}
\affil{Research Center for Nuclear Physics (RCNP), Osaka University, Ibaraki 567-0047, Japan}
\author[2]{Masaki Sasano}
\affil{Center for Nuclear Study (CNS), The University of Tokyo, Tokyo 113-0001, Japan}
\author[2]{Tomohiro Uesaka}



\date{\today}

\begin{abstract}
To discuss the dineutron correlation in the ground state, the quasi-free neutron knockout reaction on $^6$He is investigated.
In the present work, the momentum distribution of the two emitted neutrons is calculated with the $\alpha$~+~$n$~+~$n$ three-body model and a simple reaction model to assess the effects of the knockout process via the $^5$He resonance and the target dependence in the momentum distribution.
A clear signature of the dineutron correlation can be obtained by choosing the kinematical condition so the process via the $^5$He resonance is excluded because the $^5$He resonance drastically changes the momentum distribution.
Using the proton target is important in the quantitative discussion on the dineutron correlation by the knockout reaction.
In addition to theoretical arguments, a possible experimental setup to determine the dineutron correlation via the quasi-free knockout reaction is considered.
\end{abstract}


\maketitle


%
%
%
\section{introduction}
The two-neutron halo structure observed in $^6$He, $^{11}$Li, etc. is one of the most interesting topics in neutron-rich nuclei~\cite{Tanihata85,Tanihata13}.
Two-neutron halo nuclei are Borromean systems, where the binary subsystems do not have bound states.
They have been extensively studied theoretically and experimentally to understand their exotic binding mechanisms.
Theoretically, two-neutron halo nuclei have been investigated using the core~+~$n$~+~$n$ three-body models~\cite{Zhukov93,Csoto94,Suzuki91,Funada94,Aoyama95,Hagino05,Myo01,Myo08,Kikuchi10,Kikuchi13a}, which indicate that the correlations between halo neutrons are important in the bindings~\cite{Zhukov93,Aoyama95}.
It has also been shown that such a two-neutron correlation in the ground state is characterized as a spatially correlated neutron pair, the so-called dineutron~\cite{Zhukov93,Hagino05,Funada94,Myo08}.

Coulomb breakup reactions have been employed to experimentally investigate the correlation between halo neutrons through the excitation mechanism of two-neutron halo nuclei~\cite{Aumann99,Wang02,Ieki93,Shimoura95,Zinser97,Nakamura06}.
The cross sections in the Coulomb breakup reactions exhibit a low-lying peak just above the breakup thresholds.
This peak should be responsible for the properties in weakly bound halo neutrons.
The investigation of the ground-state structure of the dineutron based on a sum rule for the low-lying $E1$ transition have led to a lot of arguments concerning the relation between the sum-rule value and the geometrical configuration of halo neutrons~\cite{Kikuchi13a,Nakamura06,Esbensen92,Hagino07}.
In most cases, it is assumed that the $E1$ transition strength is directly deduced from the Coulomb breakup cross section and a two-neutron halo nucleus is described as the core~+~$n$~+~$n$ three-body system where the core nucleus is taken to be inert.
With the above assumption, the sum-rule value of the low-lying $E1$ transition depends only on the geometrical configuration of the halo neutrons in the ground state~\cite{Esbensen92}.
This sum rule is referred to as the ``cluster sum rule".
The opening angle between two halo neutrons, which is a good measure of the spatial correlation of halo neutrons, has been obtained with the cluster sum rule:
For the $^{11}$Li case, the opening angle between two halo neutrons was derived as $48^{+14}_{-18}$ degrees from the observed Coulomb breakup cross section~\cite{Nakamura06}. 

Theoretically, on the other hand, it is shown that the Coulomb breakup reactions have some problems when extracting information about the ground-state structure of two-neutron halo nuclei.
Firstly, the low-lying peaks in the cross sections are governed by strong final-state interactions (FSIs) and the sequential decay via the core~+~$n$ resonance dominates the low-lying peak~\cite{Kikuchi10,Kikuchi13a}.
This fact shows that the geometrical configuration in the ground state cannot be extracted from observables other than the cluster sum-rule value.
Secondly, the cluster sum rule itself would be invalid when the ground state of two-neutron halo nuclei contains a certain amount of the excited components of the core nucleus:
In the cluster sum rule, the core nucleus is assumed to be inert.
However, in the $^{11}$Li case, the sum rule value is reduced by about 15~\% upon considering the $^9$Li core excitation~\cite{Kikuchi13a}.
To overcome the above problems and to discuss the dineutron more quantitatively, other kinds of reactions are desired.

The knockout reaction is another useful tool to investigate the ground-state structure of nuclei.
In fact, for two-neutron halo nuclei, $^{11}$Li and $^{14}$Be, the one-neutron knockout reaction has been employed by bombarding them onto a $^9$Be target, and the contributions of the S- and P-states were discussed using the angular distributions of the emitted neutrons~\cite{Simon07}.
In the results of Ref.~\cite{Simon07}, the relative contributions of different partial waves in $^{11}$Li were determined by considering the interference between different parity partial waves, which appears as asymmetry in the angular distribution.
Although the interference between different parity waves is critical for the dineutron correlation, the dineutron correlation is not discussed in Ref.~\cite{Simon07}.
Thus, It should be of interest to discuss the dineutron correlation from the asymmetry in the angular distribution of the knocked-out neutron from the two-neutron halo nuclei.

The purpose of this work is to investigate the possibility of extracting more quantitative and more reliable information about the dineutron correlation in Borromean nuclei by means of the proton-induced neutron knockout reaction.
Although $^6$He is used as an example, the method in this paper is applicable to any Borromean nuclei.
In one-neutron knockout reactions of two-neutron halo nuclei, the other halo neutron is immediately emitted due to the Borromean nature.
We calculate the momentum distribution of the two emitted neutrons and discuss the spatial correlation between the neutrons.
In particular, the distribution of the opening angle between the momentum vectors of the two emitted neutrons is determined.
We hereafter refer this opening angle between the momentum vectors to as the ``correlation angle''.

To discuss the spatial correlations, we consider the quasi-free condition to minimize the FSIs in the knockout reaction.
In the quasi-free condition, the knocked-out neutron with a high-momentum transfer is almost free from the FSIs; hence, it provides information about the ground-state structure.
Furthermore, it is essential to use a probe as it allows the inner part of the two-neutron wave function to provide a clear signature of the dineutron correlation from the knockout reaction.
In this work, we show that a proton target is a suitable probe.
Comparing the results of proton and $^{12}$C targets shows that a proton target contrasts to heavy ions.

It is also important to minimize the effect of the knockout process via the $^5$He resonance to discuss the dineutron correlation in $^6$He.
If the knockout reaction is dominated due to the $^5$He resonance, the correlation angle distribution is governed by the $3/2^-$ single-particle orbit corresponding to the $^5$He resonance.
Such a dominance of the single spin-parity orbit suppresses the asymmetry in the distribution because the asymmetry originates from the interference between different parity single-particle orbits.
In this work, we estimate the effect of the process via the $^5$He resonance on the distribution and try to elucidate the kinematical condition that excludes the contribution from the knockout process via the $^5$He resonance.

In the last part, a possible experimental setup that meets all the requirements of the theoretical arguments is considered.
A high-momentum transfer, high statistics, and availability to pin down the state of the core are the keys.

This paper is organized as follows.
In Sec.~\ref{sec:mod}, we briefly explain the $\alpha$~+~$n$~+~$n$ three-body model for $^6$He and the calculation of the knockout reaction.
In Sec.~\ref{sec:res}, we calculate the momentum distribution and discuss the effect of the knockout process via the $^5$He resonance and target dependence in the momentum distribution.
From the calculated distribution, we show the possibility of directly measuring the dineutron in $^6$He.
In addition to the theoretical arguments, a possible experimental setup to determine the dineutron correlation via the quasi-free knockout reaction is considered.
All the results and discussions are summarized in Sec.~\ref{sec:sum}.
\section{model\label{sec:mod}}
\subsection{$\alpha$~+~$n$~+~$n$ three-body model for $^6$He}
Before discussing the knockout reaction, the $\alpha$~+~$n$~+~$n$ three-body model for $^6$He is briefly explained.
To solve the relative motion of the $\alpha$~+~$n$~+~$n$ system, we employed the orthogonality condition model (OCM)~\cite{Saito77}.

In the $\alpha$~+~$n$~+~$n$ OCM, the wave function for $^6$He is described as
\begin{equation}
\Phi_{^6\text{He}} = \Phi_\alpha \chi_{nn},
\end{equation}
where the wave function of the $\alpha$ core is expressed as $\Phi_\alpha$.
The $\alpha$ core is described by the harmonic oscillator wave function for the $(0s_{1/2})^4$ configuration.
To reproduce the observed charge radius of $^4$He, the oscillator length $b_c$ is taken as 1.4 fm.

The relative wave function of the $\alpha$~+~$n$~+~$n$ system is obtained by solving the following Schr\"odinger equation;
\begin{equation}
\hat{H} \chi_{nn} = E\chi_{nn}.
\label{eq:sch}
\end{equation}
Equation~(\ref{eq:sch}) can be accurately solved with a few-body technique.
Here we employed the variational method called the hybrid-$VT$ model whose detailed explanation is given in Ref.~\cite{Aoyama95}.

The total Hamiltonian in Eq.~(\ref{eq:sch}) is given by
\begin{equation}
\hat{H} = \sum_{i=1}^3 \hat{t}_i - \hat{T}_\text{cm} + \sum_{i=1}^2 \hat{V}_{\alpha n} (\boldsymbol{\xi}_i) + \hat{V}_{nn} + \hat{V}_\text{PF}  + \hat{V}_{\alpha n n},
\label{eq:ham}
\end{equation}
where $\hat{t}_i$ and $\hat{T}_\text{cm}$ are the kinetic operators for the $i$-th particle and the center-of-mass motion in the $\alpha$~+~$n$~+~$n$ system, respectively.
For two-body interactions of $\alpha$-$n$ and $n$-$n$, we used the effective interactions, which reproduce the scattering observables for each subsystem.
We used the KKNN potential~\cite{Kanada79} for $\hat{V}_{\alpha n}$, where $\boldsymbol{\xi}_i$ is the relative coordinates between the $\alpha$ particle and the $i$-th neutron.
For $\hat{V}_{nn}$, we used the Minnesota force~\cite{Thompson77} with an exchange parameter of 0.95.

The Pauli principle between the $\alpha$ core and the neutrons is taken into account by the so-called pseudo potential $\hat{V}_\text{PF}$.
The pseudo potential is a projection operator to remove the Pauli forbidden state $\phi_\text{PF}$ from the relative motion between the $\alpha$ core and the neutrons, and is given by
\begin{equation}
\hat{V}_\text{PF} = \lambda | \phi_\text{PF} \rangle \langle \phi_\text{PF} |.
\end{equation}
The Pauli forbidden state is defined as the $0s$ orbits occupied by the $\alpha$ core.
In the present calculation, the strength $\lambda$ is taken as $10^6$ MeV.

\begin{table}[tb]
\caption{\label{tab:gsp} Ground-state properties of $^6$He. Two-neutron separation energy $S_{2n}$ and matter and charge radii $R_\text{m}$ and $R_\text{ch}$ are presented.
Probabilities of the partial wave components $P((l_j)^2)$ in the relative coordinate set shown in Fig.~\ref{fig:coord} are also listed.}
\centering
\begin{tabular}{|c|cc|}
\hline
& Present & Exp. \\
\hline
$S_{2n}$ (MeV) & 0.975 & 0.973\footnote{Reference \cite{Tilley02}} \\
$R_\text{m}$ (fm) & 2.46 & 2.48$\pm$0.03\footnote{Reference \cite{Tanihata88}} \\
&& 2.33$\pm$0.04\footnote{Reference \cite{Tanihata92}} \\
&& 2.50\footnote{Reference \cite{Al-Khalili98}} \\
$R_\text{ch}$ (fm) & 2.04 & 2.068(11)\footnote{Reference \cite{Mueller07}} \\
\hline
$P((p_{3/2})^2)$ & 87.5 \% & \\
$P((p_{1/2})^2)$ & 3.6 \% & \\
$P((s_{1/2})^2)$ & 7.9 \% & \\
$P((d_{5/2})^2)$ & 0.4 \% & \\
$P((d_{3/2})^2)$ & 0.1 \% & \\
$P((f_{7/2})^2)$ & 0.3 \% & \\
$P((f_{5/2})^2)$ & 0.1 \% & \\
\hline
\end{tabular}
\end{table}

In the $\alpha$~+~$n$~+~$n$ OCM, the ground state of $^6$He is slightly underbound using only two-body interactions.
To reproduce the ground-state properties, we introduced the effective $\alpha nn$ three-body interaction~\cite{Myo03,Kikuchi10}, which is given by
\begin{equation}
\hat{V}_{\alpha nn} = V_3 e^{-\mu\left(\xi_1^2 + \xi_2^2\right)},
\end{equation}
where $V_3$ and $\mu$ are determined to reproduce the observed binding energy and matter radius of the $^6$He ground state.
Here $V_3$ and $\mu$ are taken as $-1.503$ MeV and $0.07/b_c^2$ fm$^{-2}$, respectively.
Table~\ref{tab:gsp} lists the ground-state properties obtained with the three-body interaction.

\subsection{One-neutron knockout cross section from $^6$He}
\begin{figure}[tb]
\centering{\includegraphics[clip,width=7.5cm]{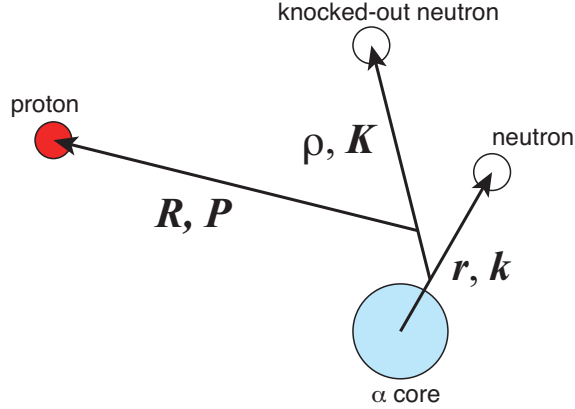}}
\caption{\label{fig:coord}
(Color online) Relative coordinates and momenta employed in the present calculation.}
\end{figure}

We considered the neutron knockout reaction on $^6$He by a proton target at 250 MeV/nucleon.
The cross section for the ($p$,$pn$) reaction on $^6$He is given by
\begin{equation}
\frac{d^9\sigma}{d\mathbf{k} d\mathbf{K} d\mathbf{P}} \propto
\frac{(2\pi)^4 \mu_R}{\hbar^2 P_0} \left|\mathcal{T}_{\mathbf{P}_0}(\mathbf{k},\mathbf{K},\mathbf{P}) \right|^2,
\label{eq:Xsec}
\end{equation}
where $P_0$ is the incident momentum of $^6$He in the center-of-mass system and $\mu_R$ is the reduced mass corresponding to the relative motion between the proton and $^6$He.
The $T$-matrix in Eq.~(\ref{eq:Xsec}) is defined as
\begin{equation}
\mathcal{T}_{\mathbf{P}_0}(\mathbf{k},\mathbf{K},\mathbf{P})
= \Big\langle \phi_0(\mathbf{P}, \mathbf{R}) \otimes \Psi_{^6\text{He}}(\mathbf{k},\mathbf{K},\mathbf{r},\boldsymbol{\rho}) \Big| \hat{V}_f \Big| \psi(\mathbf{P}_0,\mathbf{R}) \otimes \Phi_\text{gs}(\mathbf{r},\boldsymbol{\rho})\Big\rangle
\label{eq:Tmat}
\end{equation}
in the post-form representation, where $\phi_0$ is a plane wave between the target and the projectile.
The interaction in the final state $\hat{V}_f$ is the sum of the interactions between the proton and each constituent particle in $^6$He.
The wave functions of $^6$He in the initial ground state and the final scattering one are represented by $\Phi_\text{gs}$ and $\Psi_\text{$^6$He}$, respectively.
The initial scattering wave between the target and the projectile is given by $\psi$.
The relative coordinates and momenta in Eqs.~(\ref{eq:Xsec}) and (\ref{eq:Tmat}) are defined as shown in Fig.~\ref{fig:coord}.

A simple reaction model where the knockout process via the $^5$He resonance and the absorption effect by the target nucleus are taken into account is used in the calculation.
We approximate the initial scattering wave $\psi$ as
\begin{equation}
\psi(\mathbf{P}_0,\mathbf{R}) = D(\mathbf{R}) e^{i\mathbf{P}_0\cdot\mathbf{R}},
\end{equation}
where $D(\mathbf{R})$ is the damping factor for the absorption by the proton target.
Furthermore, only the $p$-$n$ interaction $\hat{V}_{p\text{-}n}$ is considered as a residual interaction for the neutron knockout by a proton target.
Hence, the other interactions in $\hat{V}_f$ are operated on the wave function in the final state $\phi_0$.
Operating these interactions on the final-state wave function transforms $\phi_0$ to
\begin{equation}
\phi_0(\mathbf{P},\mathbf{R}) \to \psi'(\mathbf{P}, \mathbf{R}) = D'(\mathbf{R}) e^{i\mathbf{P}\cdot\mathbf{R}},
\end{equation}
where $D'(\mathbf{R})$ is the damping factor coming from the final-state interactions between the $\alpha$ core and the outgoing particles.

From the above procedure, the $T$-matrix in Eq.~(\ref{eq:Tmat}) can be rewritten as
\begin{equation}
\mathcal{T}_{\mathbf{P}_0}(\mathbf{k},\mathbf{K},\mathbf{P})
= \Big\langle D'(\mathbf{R}) e^{i\mathbf{P}\cdot\mathbf{R}} \otimes \Psi_{^6\text{He}}(\mathbf{k},\mathbf{K},\mathbf{r},\boldsymbol{\rho}) \Big| \hat{V}_{p\text{-}n} \Big| D(\mathbf{R}) e^{i\mathbf{P}_0\cdot\mathbf{R}} \otimes \Phi_\text{gs}(\mathbf{r},\boldsymbol{\rho})\Big\rangle.
\label{eq:Tmat'}
\end{equation}
To calculate the $T$-matrix in Eq.~(\ref{eq:Tmat'}), we employed the zero-range interaction for $V_{p\text{-}n}$.
The $T$-matrix using the zero-range interaction gives a similar result to the calculation with an averaged $NN$ cross section over kinematics, while the absolute value of the knockout cross section cannot be reproduced.

When the zero-range interaction is employed, the $T$-matrix in Eq.~(\ref{eq:Tmat'}) is obtained as
\begin{equation}
\mathcal{T}_{\mathbf{P}_0}(\mathbf{k},\mathbf{K},\mathbf{P})
= V_0 \Big\langle \Psi_{^6\text{He}}(\mathbf{k},\mathbf{K},\mathbf{r},\boldsymbol{\rho}) \Big| \tilde{D}(\boldsymbol{\rho})  e^{i\mathbf{q}\cdot\boldsymbol{\rho}}  \Big| \Phi_\text{gs}(\mathbf{r},\boldsymbol{\rho})\Big\rangle,
\label{eq:TPWIA}
\end{equation}
where $V_0$ is the strength of the $p$-$n$ interaction and $\mathbf{q}$ is related to the transferred momentum as $\mathbf{q} = 5(\mathbf{P}_0 - \mathbf{P})/6$.
The damping factor $\tilde{D}(\boldsymbol{\rho})$ in Eq.~(\ref{eq:TPWIA}) is given by
\begin{equation}
\tilde{D}(\boldsymbol{\rho}) = D\left(\frac{5}{6}\boldsymbol{\rho}\right) \cdot D'\left(\frac{5}{6}\boldsymbol{\rho}\right).
\label{eq:Dfac}
\end{equation}

To estimate the absorption effect on the $T$-matrix, we defined the damping factor $\tilde{D}$ by the product of the eikonal $S$-matrices for the target and the two outgoing particles
\begin{equation}
\tilde{D}(b) = \prod_{i=1}^3 \exp\left[\frac{1}{i\hbar v_i} \int_{-\infty}^\infty U_i(b,z) dz\right],
\end{equation}
where $i$ is the index of the scattering particles, $v_i$ is the velocity of the $i$th particle relative to the $\alpha$ particle, and $U_i$ is the distorting potential for the $i$th particle.
The damping factor $\tilde{D}$ is obtained as a function of the impact parameter $b$.
Because we strive to roughly estimate the absorption effect, we assumed that the $b$ dependence of $\tilde{D}$ is similar to its $\rho$ dependence.

For the nucleon-nucleus distorting potential, we adopted the Dirac phenomenology~\cite{Cooper09}; for $\alpha$-$^{12}$C, the nucleon-$^{12}$C potential is folded by the density of the $\alpha$ particle following the nucleon-nucleus folding model calculation in Ref.~\cite{Egashira14}.
It is noted that the distorting potentials are given by finite-range ones and only the imaginary part of each potential is included in the evaluation of the damping factor $\tilde{D}$.
A more proper treatment of the distortion effects, including the distortion by the real parts of $U_i$, will be discussed in the forthcoming paper.

Considering the quasi-free condition, the scattering wave function of $^6$He, $\Psi_\text{$^6$He}$, is described as
\begin{equation}
\Psi_\text{$^6$He}(\mathbf{k},\mathbf{K},\mathbf{r},\boldsymbol{\rho})
= \psi_{\alpha\text{-}n} (\mathbf{k},\mathbf{r}) \otimes e^{i\mathbf{K}\cdot\boldsymbol{\rho}}.
\label{eq:6HeSW}
\end{equation}
The relative motion between the knocked-out neutron and the rest is described by a plane wave since the knocked-out neutron is free from the FSIs in the quasi-free condition.
To take the process via the $^5$He resonance in the knockout reaction into account, the $^5$He residue is expressed by the exact scattering wave function of $\alpha$~+~$n$ $\psi_{\alpha\text{-}n}$.
The wave function $\psi_{\alpha\text{-}n}$ is solved with the same $\alpha$-$n$ interaction as that used in Eq.~(\ref{eq:ham}).
Combining Eqs.~(\ref{eq:TPWIA}) and (\ref{eq:6HeSW}) gives
\begin{equation}
\mathcal{T}_{\mathbf{P}_0}(\mathbf{k},\mathbf{K},\mathbf{P})
= V_0 \Big\langle \psi_{\alpha\text{-}n} (\mathbf{k},\mathbf{r}) \otimes e^{i\mathbf{K}\cdot\boldsymbol{\rho}} \Big| \tilde{D}(\boldsymbol{\rho}) e^{i\mathbf{q}\cdot\boldsymbol{\rho}}  \Big| \Phi_\text{gs}(\mathbf{r},\boldsymbol{\rho})\Big\rangle.
\label{eq:TmatP}
\end{equation}
From Eq.~(\ref{eq:TmatP}), we also defined the following $T$-matrix;
\begin{equation}
\begin{split}
\mathcal{T}(\mathbf{k},\mathbf{K}') &= \iint d\mathbf{K} d\mathbf{P} \mathcal{T}_{P_0} (\mathbf{k},\mathbf{K},\mathbf{P})\delta \left(\mathbf{K}'-\mathbf{K}+\mathbf{q}\right)\\
&= V_0 \Big\langle \psi_{\alpha\text{-}n} (\mathbf{k},\mathbf{r}) \otimes e^{i\mathbf{K}'\cdot\boldsymbol{\rho}}  \Big| \tilde{D}(\boldsymbol{\rho}) \Big| \Phi_\text{gs}(\mathbf{r},\boldsymbol{\rho})\Big\rangle,
\end{split}
\label{eq:TmatPW}
\end{equation}
where $\mathbf{K}' = \mathbf{K}-\mathbf{q}$.

Using Eqs.~(\ref{eq:Xsec}) and (\ref{eq:TmatPW}) the double-differential cross section is expressed as
\begin{equation}
\frac{d^2\sigma}{dk_{\alpha\text{-}n}d\theta}
\propto \frac{(2\pi)^4 \mu_R}{\hbar^2 P_0} \cdot V_0^2 \cdot \frac{d^2W}{dk_{\alpha\text{-}n}d\theta},
\end{equation}
where $k_{\alpha\text{-}n}$ is the relative momentum between the $\alpha$ particle and the neutron in the $^5$He residue, and $\theta$ is the correlation angle between two momenta $\mathbf{k}$ and $\mathbf{K}$.
In the present analysis, we calculated $d^2W/dk_{\alpha\text{-}n}d\theta$ to discuss the spatial correlation between halo neutrons in $^6$He.
The two-dimensional momentum distribution $d^2W/dk_{\alpha\text{-}n}d\theta$ is defined as
\begin{equation}
\frac{d^2W}{dk_{\alpha\text{-}n}d\theta}
= \iint d\mathbf{k} d\mathbf{K}' \delta\left(k-k_{\alpha\text{-}n}\right) \delta\left(\theta'-\theta\right) \left| \Big\langle \psi_{\alpha\text{-}n} (\mathbf{k},\mathbf{r}) \otimes e^{i\mathbf{K}'\cdot\boldsymbol{\rho}}  \Big| \tilde{D}(\boldsymbol{\rho}) \Big| \Phi_\text{gs}(\mathbf{r},\boldsymbol{\rho})\Big\rangle \right|^2,
\label{eq:WDD}
\end{equation}
where $\cos{\theta'} = \hat{\mathbf{k}}\cdot\hat{\mathbf{K}}'$.
This distribution gives
\begin{equation}
\iint \frac{d^2W}{dk_{\alpha\text{-}n}d\theta} dk_{\alpha\text{-}n} d\theta = 1
\end{equation}
when $\tilde{D}(\boldsymbol{\rho}) = 1$.
We also defined the correlation angle distribution for two emitted neutrons by integrating Eq.~(\ref{eq:WDD}) over $k_{\alpha\text{-}n}$ as
\begin{equation}
\frac{dW}{d\theta} = \int \frac{d^2W}{dk_{\alpha\text{-}n}d\theta} dk_{\alpha\text{-}n}.
\label{eq:WD}
\end{equation}
\section{results\label{sec:res}}
\subsection{Dineutron correlation in the ground state}
Before discussing the knockout reaction, we first demonstrated the momentum distribution of the two halo neutrons in the ground state of $^6$He.
To obtain the ground-state momentum distribution, we calculated $d^2W/dk_{\alpha\text{-}n}d\theta$ by neglecting the absorption effect and the $\alpha$-$n$ interaction in the final state.
Thus,  we assumed that the damping factor $\tilde{D}$ is unity and replaced the scattering wave function of $\alpha$~+~$n$ $\psi_{\alpha\text{-}n} (\mathbf{k},\mathbf{r})$ with the plane wave $e^{i\mathbf{k}\cdot\mathbf{r}}$ in Eq.~(\ref{eq:TmatPW}).

\begin{figure}[tb]
\centering{\includegraphics[clip,width=10cm]{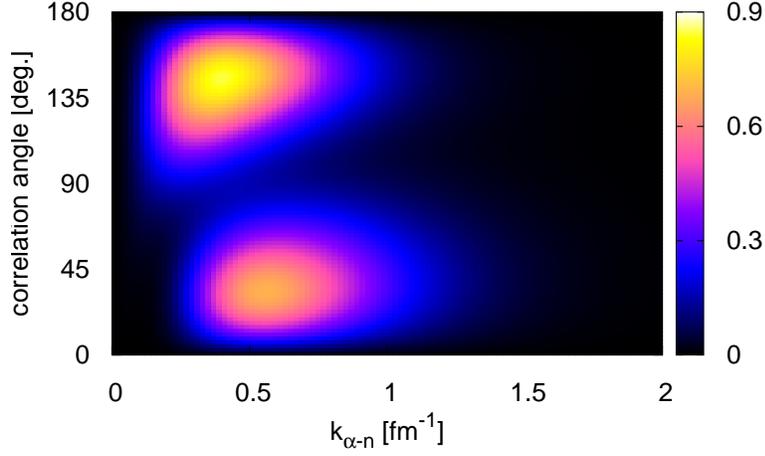}}
\caption{\label{fig:gsdis}
(Color online) Two-dimensional momentum distribution of the halo neutrons in the ground state of $^6$He, which is calculated neglecting absorption and the $\alpha$-$n$ FSI.}
\end{figure}

Figure~\ref{fig:gsdis} shows the calculated momentum distribution.
From the definition of $d^2W/dk_{\alpha\text{-}n}d\theta$ in Eq.~(\ref{eq:WDD}), the distribution in Fig.~\ref{fig:gsdis} is just the Fourier-transformed ground-state wave function of $^6$He.
The momentum distribution in Fig.~\ref{fig:gsdis} has a two-peak structure and the strength near $(k_{\alpha\text{-}n}, \theta)=(0.4$ fm$^{-1}, 135$ degrees$)$ is enhanced compared to the other peaks.
This enhancement is due to the dineutron correlation in the ground state of $^6$He, because it is intuitively understood that the angular distribution in the momentum space is opposite to that in the coordinate space.
\begin{figure}[tb]
\centering{\includegraphics[clip,width=9cm]{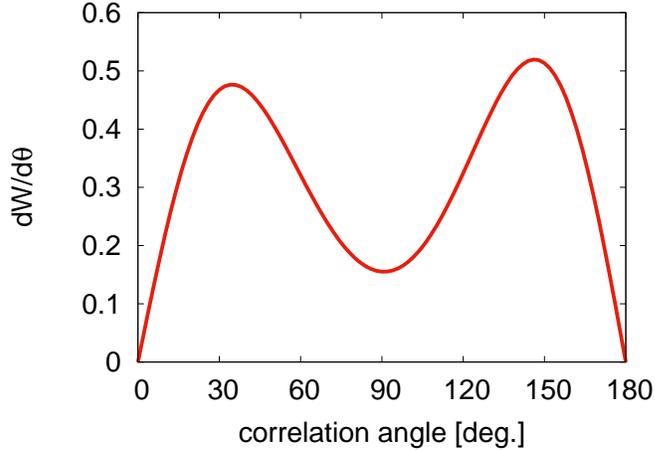}}
\caption{\label{fig:gsang}
(Color online) Correlation angle distribution of the halo neutrons, which is calculated neglecting absorption and the $\alpha$-$n$ FSI.}
\end{figure}

This dineutron correlation in the momentum space is also seen in the correlation angle distribution.
The distribution $dW/d\theta$ calculated by Eq.~(\ref{eq:WD}) is similar to the result in Fig.~\ref{fig:gsdis}.
Figure~\ref{fig:gsang} shows the correlation angle distribution, which has a two-peak structure as expected from the result in Fig.~\ref{fig:gsdis}.
The peak at $\theta > 90$ degrees, which corresponds to the dineutron correlation, is slightly higher than that at $\theta < 90$ degrees.

\begin{figure}[tb]
\centering{\includegraphics[clip,width=9cm]{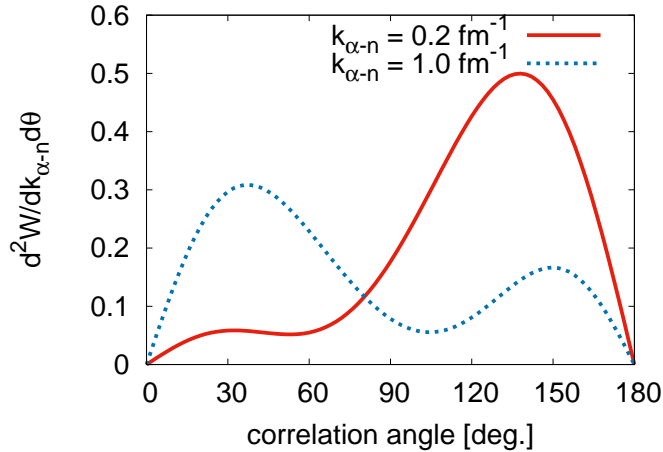}}
\caption{\label{fig:gsangfix}
(Color online) Ground-state momentum distributions of the halo neutrons with fixed $k_{\alpha\text{-}n}$. Red (solid) and blue (dotted) lines represent the distributions for $k_{\alpha\text{-}n} = 0.2$ and 1.0 fm$^{-1}$, respectively.}
\end{figure}

The dineutron correlation seen in the distributions of Figs.~\ref{fig:gsdis} and \ref{fig:gsang} is clearer in the distribution of $d^2W/dk_{\alpha\text{-}n}d\theta$ with a fixed relative momentum for $\alpha$~+~$n$. 
Figure ~\ref{fig:gsangfix} shows the distributions with fixed values of $k_{\alpha\text{-}n}$.
Here we chose two momenta: $k_{\alpha\text{-}n}$ = 0.2 and 1.0 fm$^{-1}$.
At $k_{\alpha\text{-}n} = 0.2$ fm$^{-1}$, a significant enhancement is found at larger angles, which corresponds to the dineutron correlation.
On the other hand, the distribution at $k_{\alpha\text{-}n} = 1.0$ fm$^{-1}$ shows an enhancement at smaller correlation angle, which corresponds to the cigar-like configuration.
These results show that the dineutron correlation in the ground state is developed in the surface region of $^6$He, which corresponds to the small momentum between the $\alpha$ core and the neutron.

In our calculation, the asymmetry in the distributions in Fig.~\ref{fig:gsangfix} is predominantly due to the interference between the $(0p_{3/2})^2$ and $(1s_{1/2})^2$ components.
In the ground state of $^6$He, the mixing of the $(0p_{3/2})^2$ and $(1s_{1/2})^2$ components is caused by the strong two-neutron correlation.
The relative phase between $(0p_{3/2})^2$ and $(1s_{1/2})^2$ changes at $k_{\alpha\text{-}n} \sim 0.5$ fm$^{-1}$ because the single-particle wave function of $1s_{1/2}$ has a node at $k_{\alpha\text{-}n} \sim 0.5$ fm$^{-1}$.
This change in the relative phase results in the interference patterns at $k_{\alpha\text{-}n} =$ 0.2 and 1.0 fm$^{-1}$ to differ from each other.

\subsection{Effect of the knockout process via the $^5$He resonance}
Next, we investigated the effect of the process via the $^5$He resonance on the momentum distribution.
To investigate the effect of the process via the $^5$He resonance, we calculated the distribution by including the $\alpha$-$n$ interaction in the final states.
Here, we focused on the effect of the process via the $^5$He resonance, while the absorption effect is neglected by assuming that the damping factor in Eq.~(\ref{eq:TmatPW}) is unity. 

\begin{figure}[tb]
\centering{\includegraphics[clip,width=10cm]{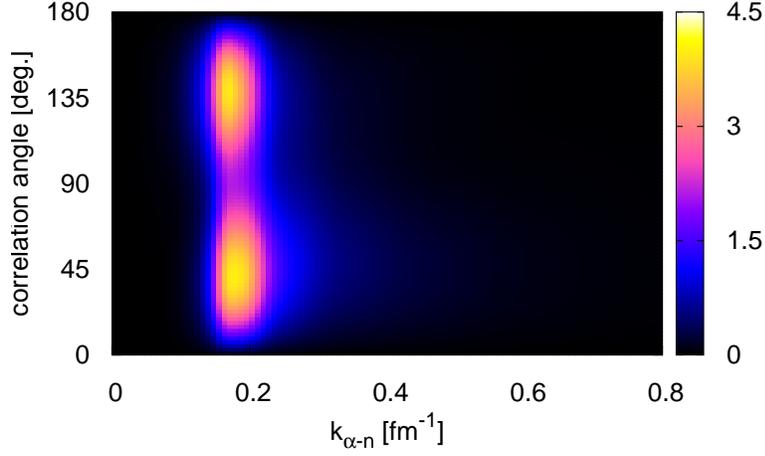}}
\caption{\label{fig:KOdis}
(Color online) Two-dimensional momentum distribution of the emitted neutrons in the knockout reaction, which is calculated without absorption from a target.}
\end{figure}

Figure~\ref{fig:KOdis} shows the calculated momentum distribution $d^2W/dk_{\alpha\text{-}n}d\theta$.
The inclusion of the $\alpha$-$n$ FSI drastically changes the distribution compared to the result in Fig.~\ref{fig:gsdis}.
The distribution is concentrated on the $k_{\alpha\text{-}n} \sim 0.2$ fm$^{-1}$ region and has a symmetric two-peak structure with respect to $\theta = 90$ degrees.
This distribution shape shows that the knockout reaction of $^6$He is dominated by the process via the $^5$He($3/2^-$) resonance.
The resonance energy of $^5$He($3/2^-$) is coincident with $k_{\alpha\text{-}n} \sim 0.2$ fm$^{-1}$ and the symmetric two-peak structure comes from the $(p_{3/2})^2$ configuration of the two emitted neutrons, which is favored by the process via the $^5$He resonance.

\begin{figure}[tb]
\centering{\includegraphics[clip,width=9cm]{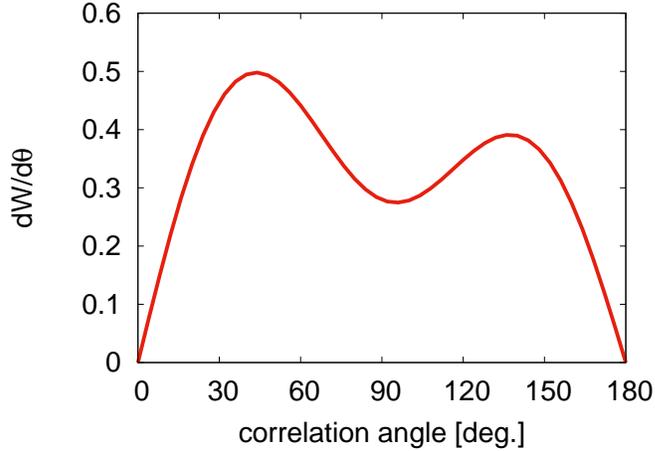}}
\caption{\label{fig:KOang}
(Color online) Correlation angle distribution of the emitted neutrons in the knockout reaction, which is calculated without absorption from a target.}
\end{figure}
Furthermore, in the correlation angle distribution $dW/d\theta$ (Fig.~\ref{fig:KOang}), any signature of the enhancement of the dineutron correlation is not found.
By including the process via the $^5$He resonance, the distribution changes so that the largest peak is at $\theta < 90$ degrees and the contribution from the dineutron correlation becomes smaller than that from the cigar-like one.

The results in Figs.~\ref{fig:KOdis} and \ref{fig:KOang} show that the inclusion of the $^5$He resonance drastically changes the distribution and the signature of the dineutron correlation from the knockout reaction of $^6$He.
To clarify the signature of the dineutron correlation in the knockout reaction of $^6$He, it is important to minimize the effect of the process via the $^5$He resonance on the momentum distribution.

\begin{figure}[tb]
\centering{\includegraphics[clip,width=9cm]{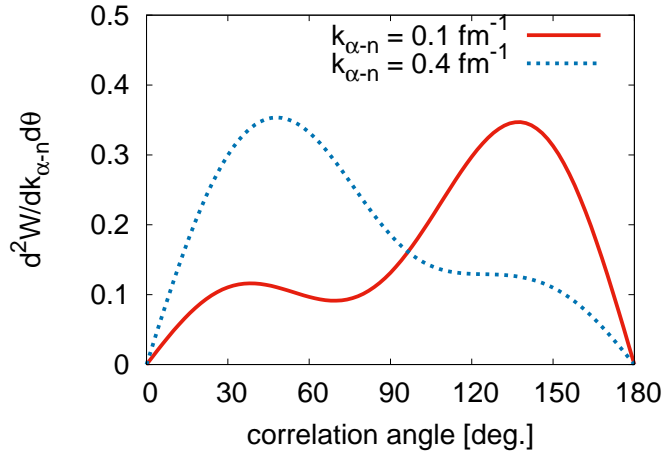}}
\caption{\label{fig:KOangfix}
(Color online) Momentum distributions of the emitted neutrons in the knockout reaction with fixed $k_{\alpha\text{-}n}$. Red (solid) and blue (dotted) lines represent the distributions for $k_{\alpha\text{-}n} = 0.1$ and 0.4 fm$^{-1}$, respectively.}
\end{figure}

For this purpose, we calculated the distribution $d^2W/dk_{\alpha\text{-}n}d\theta$ by fixing the relative momentum between the $\alpha$~+~$n$.
Here, we chose off-resonant $\alpha$-$n$ relative momenta.
The angular distributions calculated with $k_{\alpha\text{-}n} =$ 0.1 and 0.4 fm$^{-1}$, which are respectively smaller and larger than the $\alpha$-$n$ relative momentum corresponding to $^5$He(3/2$^-$) resonance, are shown in Fig.~\ref{fig:KOangfix}.
For the smaller momentum, $k_{\alpha\text{-}n} = 0.1$ fm$^{-1}$, the distribution shows the clear signature of the dineutron correlation at $\theta > 90$ degrees, while the distribution for the larger momentum, $k_{\alpha\text{-}n}$ = 0.4 fm$^{-1}$, shows the cigar-like peak.

By gating on the $\alpha$-$n$ relative momentum, the asymmetric shapes in the momentum distributions are similar to those in Fig.~\ref{fig:gsangfix}.
This fact shows that it is essential to choose the kinematical condition to exclude the process via the $^5$He resonance when discussing the dineutron correlation by the knockout reaction.
On the other hand, the correlation angle distribution calculated by integrating over $k_{\alpha\text{-}n}$, shown in Fig.~\ref{fig:KOang}, is not useful when investigating the spatial correlation such as a dineutron.

\subsection{Effect of the absorption by the target nucleus on the momentum distribution}
\begin{figure}[tb]
\centering{\includegraphics[clip,width=9cm]{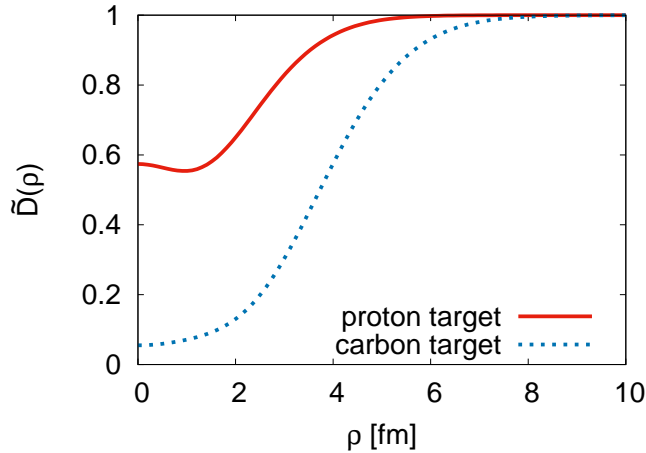}}
\caption{\label{fig:Dfac}
(Color online) Damping factor for the neutron-knockout reaction on $^6$He by a proton target at 250 MeV/nucleon, which is shown as the red (solid) line. For reference, the blue (dotted) line shows that by a $^{12}$C target.}
\end{figure}
We estimated the absorption effect by the target nucleus on the momentum distribution.
Before discussing the absorption effect, the damping factor used in the present calculation is shown.
Figure~\ref{fig:Dfac} compares the damping factor for the proton target given in Eq.~(\ref{eq:Dfac}) with that for the $^{12}$C target for reference.
The proton target is much more transparent than the $^{12}$C one.
The damping factor for the proton target is 0.57 at the origin, while that for $^{12}$C target is only 0.06.

\begin{figure}[tb]
\centering{\includegraphics[clip,width=9cm]{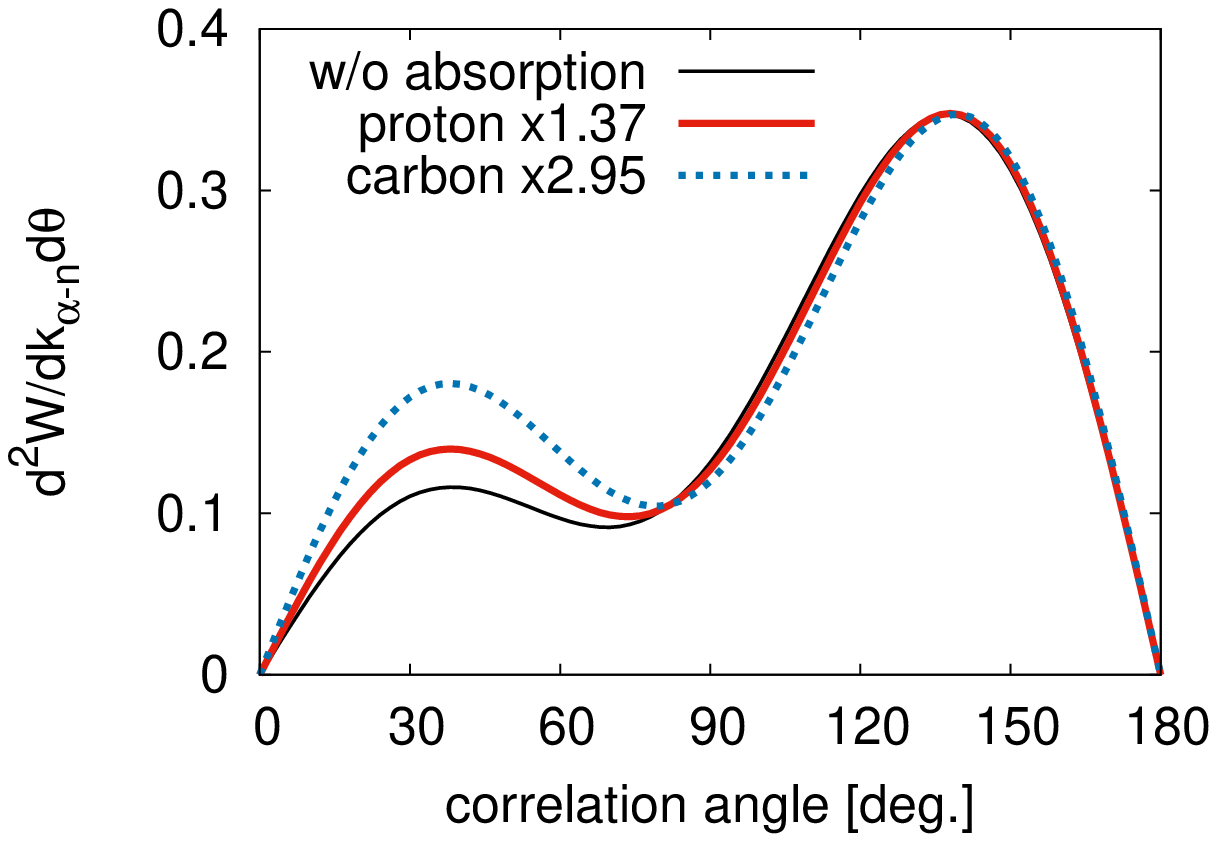}}
\caption{\label{fig:DWanglow}
(Color online) Momentum distributions of the emitted neutrons with distortion effects for $k_{\alpha\text{-}n}=0.1$ fm$^{-1}$. Red (solid) and blue (dotted) lines represent the distributions of the knockout reactions by the proton and $^{12}$C targets, respectively. Black thin line is the distribution without absorption. Distributions for the proton and carbon targets are multiplied by the factors indicated in keys to adjust the magnitude of the largest peak in the distribution without absorption.}
\centering{\includegraphics[clip,width=9cm]{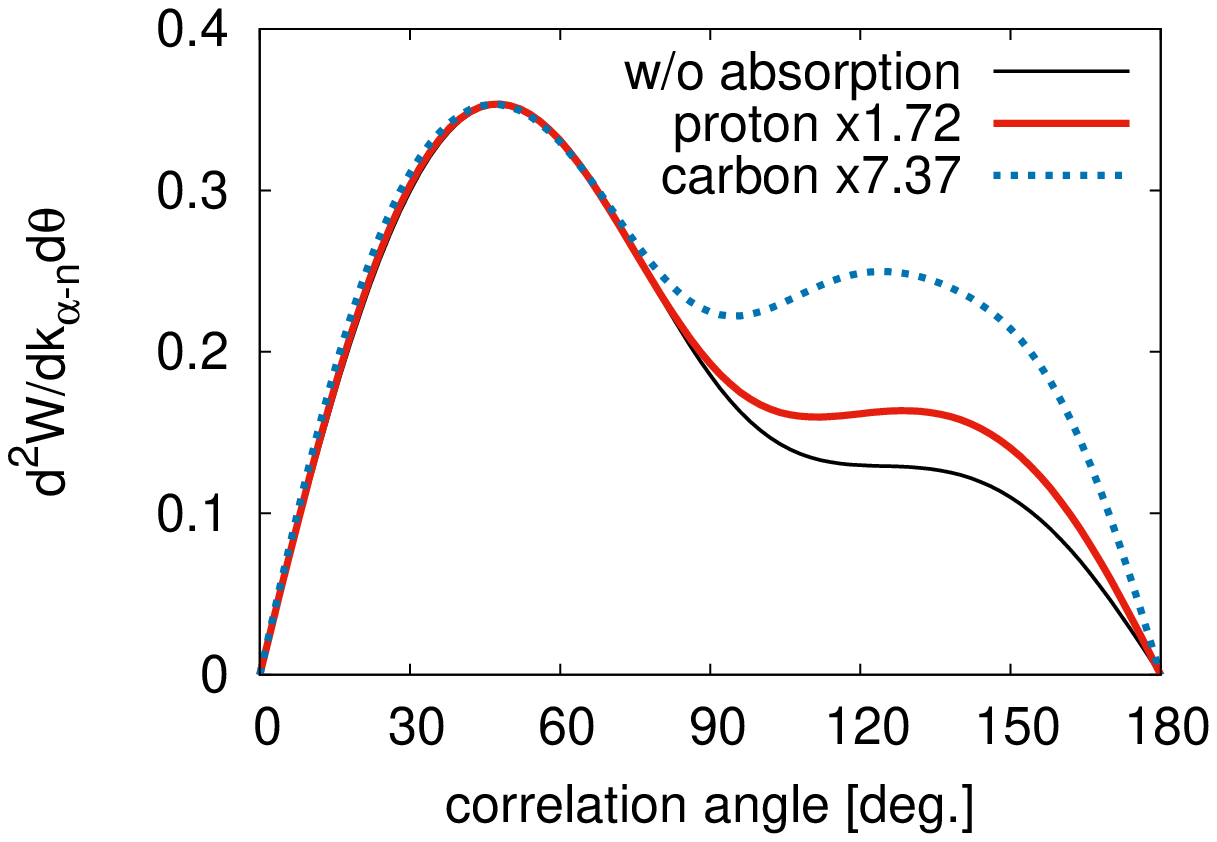}}
\caption{\label{fig:DWanghigh}
(Color online) Same as Fig.~\ref{fig:DWanglow} but for $k_{\alpha\text{-}n} = 0.4$ fm$^{-1}$.}
\end{figure}

Using the damping factors shown in Fig.~\ref{fig:Dfac}, we calculated the momentum distributions with fixed values of $k_{\alpha\text{-}n}$.
To see the target-dependence of the momentum distributions, we calculated the distributions for both the proton target and the $^{12}$C one.
Figures~\ref{fig:DWanglow} and \ref{fig:DWanghigh} show the calculated distributions for $k_{\alpha\text{-}n}=0.1$ and 0.4 fm$^{-1}$, respectively, compared to the distribution without the damping factor.

Figures~\ref{fig:DWanglow} and \ref{fig:DWanghigh} show the results where the asymmetry in the momentum distributions is suppressed as the absorption becomes stronger.
For both $k_{alpha\text{-}n} = 0.1$ and $0.4$ fm$^{-1}$, the distributions for the $^{12}$C target are more symmetric than those for the proton target, suggesting the employing the proton target is essential to investigate the spatial correlation of halo neutrons from the momentum distribution in the quasi-free knockout reaction.

Furthermore, from an experimental point of view, it is crucial to use a transparent target such as a proton.
Absorption by the target nucleus reduces the strength of the momentum distribution, and the absorption effect on the distribution is significant, especially in the case of the $^{12}$C target.
For $k_{\alpha\text{-}n} = 0.4$~fm$^{-1}$, the strength at the peak position for $^{12}$C is 7.37 times smaller than the case without absorption, while that for the proton is 1.72 times smaller.
The transparent target provides clean data, while the heavier target with a large absorption makes the experimental statistics low.

\subsection{Experimental approaches to the two-neutron correlation}
  
Here, the experimental feasibility to extract the signature of the two-neutron correlation is discussed.
First we estimated the cross section for the $(p,pn)$ reaction on $^6$He.
The cross section is calculated by the distorted-wave impulse approximation (DWIA) since we could not reproduce the absolute value of the cross section using the present model with a zero-range $p$-$n$ interaction.
In the DWIA calculation, we used the optical potentials from the Dirac phenomenology and considered the kinematical conditions listed in Table~\ref{tab:kin}.
Integrating over the kinematics, the cross section for the $(p,pn)$ reaction on $^6$He is obtained as $2\times10^2$ $\mu$b, which is subsequently used to estimate the typical value of the cross section by gating on the relative momentum $k_{\alpha\text{-}n}$.
With a gate width of $50$ keV for the $\alpha$-$n$ relative energy, the cross section is on order of 1$\mu$b.
\begin{table}[tb]
\caption{\label{tab:kin} Kinematical conditions in the inverse kinematics. The scattering and azimuthal angles for the recoiled proton and the knocked-out neutron are given in unit of degrees.}
\centering
\begin{tabular}{|cc|c|}
\hline
$^6$He projectie & incident energy & 150 MeV/nucleon \\
\hline
recoiled proton & energy acceptance & $E_p \le 50$ MeV \\
& scattering angle & $25 \le \theta_p \le 65$ \\
& azimuthal angle & $-20 \le \phi_p \le 20$ \\
\hline
knocked-out neutron & scattering angle & $25 \le \theta_n \le 65$ \\
& azimuthal angle & $160 \le \phi_n \le 200$ \\
\hline
\end{tabular}
\end{table}

The experimental extraction of the signature of the two-neutron correlation discussed above demands 
(i) selection of the $(p,pn)$ events with a sufficiently large momentum transfer, 
(ii) high statistics to divide the correlation angle distribution according to $k_{\alpha\text{-}n}$, and
(iii) a kinematically complete measurement to tag contribution from the excited core states.
In the following, an inverse kinematics experiment with a hydrogen target is considered.
A recoil proton and a knocked-out neutron could be detected in dedicated detectors placed sideways.
A decay neutron and a residual ion after the $(p,pn)$ reaction could be analyzed with a large acceptance magnetic spectrometer (e.g., the SAMURAI~\cite{Kobayashi13} at RIBF and the ALADIN at GSI) equipped with high-efficiency neutron detector arrays. 

To suppress possible effects of three-body FSIs, momenta of the recoil proton and the knocked-out neutron should be sufficiently larger than the nucleon Fermi momentum in a nucleus, which is about 1.3~fm$^{-1}$.
This condition could be fulfilled using beams with incident energies higher than 200~MeV/nucleon and restricting the recoil proton energy so that it exceeds 80~MeV.
Radioactive nuclear beams at more than 200~MeV/nucleon are presently available at the RI Beam Factory in Japan and the GSI in Germany and will be available at the FRIB in the USA in the near future.

Selecting the high-momentum transfer condition is inevitably accompanied with a small cross section.
The $(p,pn)$ cross section is estimated to be on the order of 1 $\mu$b by the DWIA calculation with a gate of 50 keV for $k_{\alpha\text{-}n}$.
To obtain sufficient statistics for the momentum distribution at each $k_{\alpha\text{-}n}$ for the small cross section, the experiment demands unprecedentedly high luminosity, typically of $10^{29}$~cm$^{-2}$sec$^{-1}$.
This could be realized with high-intensity radioactive nuclear beams at an intensity of $10^{5-6}$~sec$^{-1}$ and a thick liquid hydrogen target (thickness of  $10^{23-24}$~cm$^{-2}$).
An example of the latter is the MINOS\cite{Obertelli14}, which is a device consisting of a very thick liquid hydrogen target and a time projection chamber surrounding the target.
The target thickness is adjustable and can be as high as $10^{24}$~cm$^{-2}$.
Determination of the vertex position with the time projection chamber allows the energy loss of the beam and the residual ions in the target to be corrected.
Consequently, high luminosity experiments could be realized without a significant loss in the experimental resolutions.

A high luminosity enables tagging of the excited core contribution in Borromean nuclei.
The effects of the unbound core states, which are considered to be significant in $^{11}$Li, could be explored with invariant mass spectroscopy of the excited core states.
On the other hand, those of the bound excited core states, which is more important in heavier Borromean nuclei, could be tagged through $\gamma$-ray detection with a high-efficiency scintillator array.
New data on the effects of excited cores should reveal unexplored aspects of Borromean nuclei, resulting in a better understanding of the two-neutron correlation in the nuclei.

\section{summary\label{sec:sum}}
We investigated the dineutron correlation in $^6$He using the quasi-free neutron knockout reaction.
In the present analysis, we calculated the momentum distribution of the two emitted neutrons, and compared them with that in the ground state to discuss the spatial correlation between halo neutrons.
The pattern of the calculated momentum distribution differ from that in the ground state due to the knockout process via the $^5$He(3/2$^-$) resonance.
To minimize the effect of the process via the $^5$He resonance on the momentum distribution, we also demonstrated the momentum distribution by choosing the off-resonance region for the $\alpha$-$n$ relative momentum.
By selecting the $\alpha$-$n$ relative momentum such that the process via the $^5$He resonance is excluded, the distribution shows patterns similar to the ground state.
For a lower $\alpha$-$n$ relative momentum, the calculated distribution shows a clear enhancement at a larger correlation angle between the momenta of the two emitted neutrons.
This result suggests that the spatially correlated neutron pair is in the surface region of the halo nucleus $^6$He.
In addition, we discussed the effect of the absorption by the target nucleus on the momentum distribution.
The absorption by the target reduces not only the magnitude of the momentum distribution but also the asymmetry in the distribution.
Our results indicate that the use of the proton target, which is the most transparent probe sensitive to neutrons, is essential to investigate the spatial correlation of halo neutrons.
Furthermore, we also discussed a possible experimental setup to meet all the requirements according to the theoretical arguments.

In the present calculation, we use the simple reaction model, in which the zero-range $p$-$n$ interaction is adopted for the knockout part and only the absorption effect is taken into account in the distorting potentials.
A more quantitative analysis based on the distorted-wave impulse approximation will be performed in the forthcoming paper.
\section*{Acknowledgments}
This work was supported by a Grant-in-Aid from the Japan Society for the Promotion of Science (No. 25400255).


\begin{thebibliography}{99}
\bibitem{Tanihata85} I.~Tanihata, H.~Hamagaki, O.~Hashimoto, Y.~Shida, N.~Yoshikawa, K.~ Sugimoto, O.~Yamakawa, T.~Kobayashi, and N.~Takahashi, Phys. Rev. Lett. {\bf 55}, 2676 (1985).
\bibitem{Tanihata13} I.~Tanihata, H.~Savajols, and R.~Kanungo, Prog. Part. Nucl. Phys. {\bf 68}, 215 (2013).
\bibitem{Zhukov93} M.~V.~Zhukov, B.~V.~Danilin, D.~V.~Fedorov, J.~M.~Bang, I.~J.~Thompson, and J.~S.~Vaagen, Phys. Rep. {\bf 231}, 151 (1993).
\bibitem{Csoto94} A.~Cs\'ot\'o, Phys. Rev. C {\bf 49}, 3035 (1994).
\bibitem{Suzuki91} Y.~Suzuki, Nucl. Phys. A {\bf 528}, 395 (1991).
\bibitem{Funada94} S.~Funada, H.~Kameyama, and Y.~Sakuragi, Nucl. Phys. A {\bf 575}, 93 (1994).
\bibitem{Aoyama95} S.~Aoyama, S.~Mukai, K.~Kat\=o, and K.~Ikeda, Prog. Theor. Phys. {\bf 93}, 99 (1995).
\bibitem{Hagino05} K.~Hagino and H.~Sagawa, Phys. Rev. C {\bf 72}, 044321 (2005).
\bibitem{Myo01} T.~Myo, K.~Kat\=o, S.~Aoyama, and K.~Ikeda, Phys. Rev. C {\bf 63}, 054313 (2001).
\bibitem{Myo08} T.~Myo, Y.~Kikuchi, K.~Kat\=o, H.~Toki, and K.~Ikeda, Prog. Theor. Phys. {\bf 119}, 561 (2008).
\bibitem{Kikuchi10} Y.~Kikuchi, K.~Kat\=o, T.~Myo, M.~Takashina, and K.~Ikeda, Phys. Rev. C {\bf 81}, 044308 (2010).
\bibitem{Kikuchi13a} Y.~Kikuchi, T.~Myo, K.~Kat\=o, and K.~Ikeda, Phys. Rev. C {\bf 87}, 034606 (2013).
\bibitem{Aumann99} T.~Aumann, D.~Aleksandrov, L.~Axelsson, T.~Baumann, M.~J.~G.~Borge, L.~V.~Chulkov, J.~Cub, W.~Dostal, B.~Eberlein, T.~W.~Elze, {\it et al.}, Phys. Rev. C {\bf 59}, 1252 (1999).
\bibitem{Wang02} J.~Wang, A.~Galonsky, J.~J.~Kruse, E.~Tryggestad, R.~H.~White-Stevens, P~.D.~Zecher, Y.~Iwata, K.~Ieki, A.~Horv\'ath, F.~De\'ak, {\it et al.}, Phys. Rev. C {\bf 65}, 034306 (2002).
\bibitem{Ieki93} K.~Ieki, D.~Sackett, A.~Galonsky, C.~A.~Bertulani, J.~J.~Kruse, W.~G.~Lynch, D.~J.~Morrissey, N.~A.~Orr, H.~Schulz, B.~M.~Sherrill, {\it et al.}, Phys. Rev. Lett. {\bf 70}, 730 (1993).
\bibitem{Shimoura95} S.~Shimoura, T.~Nakamura, M.~Ishihara, N.~Inabe, T.~Kobayashi, T.~Kubo, R.~H.~Siemssen, I.~Tanihata, and Y.~Watanabe, Phys. Lett. B {\bf 348}, 29 (1995).
\bibitem{Zinser97} M.~Zinser, F.~Humbert, T.~Nilsson, W.~Schwab, H.~Simon, T.~Aumann, M.~J.~G.~Borge, L.~V.~Chulkov, J.~Cub, T.~W.~Elze, {\it et al.}, Nucl. Phys. A {\bf 619}, 151 (1997).
\bibitem{Nakamura06} T.~Nakamura, A.~M.~Vinodkumar, T.~Sugimoto, N.~Aoi, H.~Baba, D.~Bazin, N.~Fukuda, T.~Gomi, H.~Hasegawa, N.~Imai, {\it et al.}, Phys. Rev. Lett. {\bf 96}, 252502 (2006).
\bibitem{Esbensen92} H.~Esbensen and G.~F.~Bertsch, Nucl. Phys. A {\bf 542}, 310 (1992).
\bibitem{Hagino07} K.~Hagino and H.~Sagawa, Phys. Rev. C {\bf 76}, 047302 (2007).
\bibitem{Simon07} H.~Simon, M.~Meister, T.~Aumann, M.~Borge, L.~Chulkov, U.~D.~Pramanik, T.~Elze, H.~Emling, C.~Forss\'en, H.~Geissel, {\it et al.}, Nuclear Physics A {\bf 791}, 267 (2007).
\bibitem{Saito77} S. Saito, Prog. Theor. Phys. Suppl. {\bf 62}, 11 (1977).
\bibitem{Kanada79} H.~Kanada, T.~Kaneko, S.~Nagata, and M.~Nomoto, Prog. Theor. Phys. {\bf 61}, 1327 (1979).
\bibitem{Thompson77} D.~R.~Thompson, M.~Lemere, and Y.~C.~Tang, Nucl. Phys. A {\bf 286}, 53 (1977).
\bibitem{Tilley02} D.~R.~Tilley, C.~M.~Cheves, J.~L.~Godwin, G.~M.~Hale, H.~M.~Hofmann, J.~H.~Kelley, C.~G.~Sheu, and H.~R.~Weller, Nucl. Phys. A {\bf 708}, 3 (2002).
\bibitem{Tanihata88} I.~Tanihata, T.~Kobayashi, O.~Yamakawa, S.~Shimoura, K.~Ekuni, K.~Sugimoto, N.~Takahashi, T.~Shimoda, and H.~Sato, Phys. Lett. B {\bf 206}, 592 (1988).
\bibitem{Tanihata92} I.~Tanihata, D.~Hirata, T.~Kobayashi, S.~Shimoura, K.~Sugimoto, and H.~Toki, Phys. Lett. B {\bf 289}, 261 (1992).
\bibitem{Al-Khalili98} J.~S.~Al-Khalili and J.~A.~Tostevin, Phys. Rev. C {\bf 57}, 1846 (1998).
\bibitem{Mueller07} P.~Mueller, I.~A.~Sulai, A. C.~C.~Villari, J.~A.~Alc\'antara-N\'u\~nez, R.~Alves-Cond\'e, K.~Bailey, G.~W.~F.~Drake, M.~Dubois, C.~El\'eon, G.~Gaubert, {\it et al.}, Phys. Rev. Lett. {\bf 99}, 252501 (2007).
\bibitem{Myo03} T.~Myo, S.~Aoyama, K.~Kat\=o, and K.~Ikeda, Phys. Lett. B {\bf 576}, 281 (2003).
\bibitem{Cooper09} E.~D.~Cooper, S.~Hama, and B.~C.~Clark, Phys. Rev. C {\bf 80}, 034605 (2009).
\bibitem{Egashira14} K.~Egashira, K.~Minomo, M.~Toyokawa, T.~Matsumoto, and M.~Yahiro, Phys.
Rev. C {\bf 89}, 064611 (2014).
\bibitem{Kobayashi13} T.~Kobayashi, N.~Chiga, T.~Isobe, Y.~Kondo, T.~Kubo, K.~Kusaka, T.~Motobayashi, T.~Nakamura, J.~Ohnishi, H.~Okuno, {\it et al.}, Nucl. Instr. Meth. B {\bf 317}, 294 (2013).
\bibitem{Obertelli14} A.~Obertelli, A.~Delbart, S.~Anvar, L.~Audirac, G.~Authelet, H.~Baba, B.~Bruyneel, D.~Calvet, F.~Ch\^ateau, A.~Corsi, {\it et al.}, Eur. Phys. J. A {\bf 50}, 8 (2014).
\end{thebibliography}
\end{document}